\documentclass[aps,preprintnumbers,showpacs,twocolumn,superscriptaddress]{revtex4}
\usepackage{amsmath}
\usepackage{latexsym}
\usepackage{graphicx}
\usepackage{epstopdf}

\begin{document}

%%%%%%%%%%%%%%%%%
%%%   TITLE   %%%
%%%%%%%%%%%%%%%%%

\title{Accurate Evolution of Orbiting Binary Black Holes}

%%%%%%%%%%%%%%%%%%%
%%%   AUTHORS   %%%
%%%%%%%%%%%%%%%%%%%

\author{Peter~Diener}
\affiliation{Center for Computation \& Technology,
Louisiana State University, Baton Rouge, LA 70803, USA}
\affiliation{Department of Physics and Astronomy
Louisiana State University, Baton Rouge, LA 70803, USA}

\author{Frank~Herrmann}
\affiliation{Max-Planck-Institut f\"ur Gravitationsphysik,
Albert-Einstein-Institut, Am M\"uhlenberg 1, 14476 Golm, Germany}
\affiliation{Center for Gravitational Physics \& Geometry,
Penn State University, University Park, PA 16802}

\author{Denis~Pollney}
\affiliation{Max-Planck-Institut f\"ur Gravitationsphysik,
Albert-Einstein-Institut, Am M\"uhlenberg 1, 14476 Golm, Germany}

\author{Erik~Schnetter}
\affiliation{Center for Computation \& Technology,
Louisiana State University, Baton Rouge, LA 70803, USA}
\affiliation{Max-Planck-Institut f\"ur Gravitationsphysik,
Albert-Einstein-Institut, Am M\"uhlenberg 1, 14476 Golm, Germany}

\author{Edward~Seidel}
\affiliation{Center for Computation \& Technology, 
Louisiana State University, Baton Rouge, LA 70803, USA}
\affiliation{Department of Physics and Astronomy
Louisiana State University, Baton Rouge, LA 70803, USA}
\affiliation{Max-Planck-Institut f\"ur Gravitationsphysik,
Albert-Einstein-Institut, Am M\"uhlenberg 1, 14476 Golm, Germany}

\author{Ryoji Takahashi}
\affiliation{Center for Computation \& Technology, 
Louisiana State University, Baton Rouge, LA 70803, USA}

\author{Jonathan Thornburg}
\affiliation{Max-Planck-Institut f\"ur Gravitationsphysik,
Albert-Einstein-Institut, Am M\"uhlenberg 1, 14476 Golm, Germany}

\author{Jason Ventrella}
\affiliation{Center for Computation \& Technology, 
Louisiana State University, Baton Rouge, LA 70803, USA}

%%%%%%%%%%%%%%%%
%%%   DATE   %%%
%%%%%%%%%%%%%%%%

\date{19 December 2005}

%%%%%%%%%%%%%%%%%%%%
%%%   ABSTRACT   %%%
%%%%%%%%%%%%%%%%%%%%

\begin{abstract}
We present a detailed analysis of binary black hole evolutions in the
last orbit, and demonstrate consistent and convergent results for the
trajectories of the individual bodies. The gauge choice can
significantly affect the overall accuracy of the evolution. It is
possible to reconcile certain gauge dependent discrepancies by
examining the convergence limit.  We illustrate these results using an
initial data set recently evolved by
Br\"{u}gmann \textit{et~al.} (\textit{Phys.~Rev.~Lett.}~\textbf{92}, 211101).
For our highest resolution and most
accurate gauge, we estimate the duration of this data set's last orbit
to be approximately $59 M_\textrm{ADM}$.
\end{abstract}

%%%%%%%%%%%%%%%%
%%%   PACS   %%%
%%%%%%%%%%%%%%%%


%%%%%%%%%%%%%%%%%%%%
%%%   PREPRINT   %%%
%%%%%%%%%%%%%%%%%%%%

\preprint{AEI-2005-197}
\preprint{LSU-REL-121605}

%%%%%%%%%%%%%%%%%%%%%
%%%   MAKETITLE   %%%
%%%%%%%%%%%%%%%%%%%%%

\maketitle

%%%%%%%%%%%%%%%%%%%%%%%%
%%%   INTRODUCTION   %%%
%%%%%%%%%%%%%%%%%%%%%%%%

\emph{Introduction.}
\setcounter{section}{1}
Over the course of the next decade, instruments capable of detecting
gravitational radiation (such as LIGO, VIRGO, TAMA, GEO600) are
expected to open a new observational window on the universe.  The
collision of binary compact objects such as black holes (BHs) is one
of the most promising sources for first generation gravitational wave
observatories.  The theoretical framework for modelling binary BH
systems is the complete set of nonlinear Einstein equations.
Intensive efforts to develop numerical codes able to solve these
equations using supercomputers, have shown that it is now possible
to evolve BHs for periods of an orbit~\cite{Bruegmann:2003aw,
Alcubierre2003:pre-ISCO-coalescence-times, Pretorius:2005gq}.
If these simulations are to produce waveforms useful for detector
searches, high demands are placed on their
accuracy~\cite{Miller:2005qu}.

The near-zone dynamics of binary BH systems are notoriously difficult
to simulate, and to analyse.  Using the BSSN formulation and a
particular set of gauges, a series of BBH configurations,
corresponding to initial data in quasi-circular orbit at successively
larger separations~\cite{Cook94}, were all found to coalesce in
slightly more than a half
orbit~\cite{Alcubierre2003:pre-ISCO-coalescence-times}.  A similar
BSSN evolution carried out using somewhat different gauges and
numerical methods for another data set, slightly further out along the
orbital sequence, was found to evolve for much more than the estimated
orbital timescale $114 M_{\textrm{ADM}}$ without finding a
common apparent horizon (AH)~\cite{Bruegmann:2003aw}.
In fact, no common horizon was found long after the BHs would
reasonably be expected to have merged.

In this paper, we carry out an evolution of the same data set and
show that it does indeed carry out a complete orbit before a
common AH forms. As the BH separation decreases, a local measure of the
angular velocity~$\Omega$ increases, so that the duration of the final
orbit is approximately $59 M$. The trajectories are convergent for
a range of resolutions and within a class of gauge conditions.

However, we do find that very high resolutions are required in order to
obtain evolutions close to the continuum limit. The resolutions we
have applied here are significantly higher than those used in
analogous BH evolutions to date, except for~\cite{Pretorius:2005gq}
where similar resolutions were used. With insufficient resolution, we show
that it is possible to substantially under- or over-predict the orbital
period. We also find that apparently small deviations in the chosen
coordinate conditions (gauge), based on choice of parameters within a
particular family, can have a strong influence on the discretisation
error within the subsequent evolution.

%%%%%%%%%%%%%%%%%%%%%%%%
%%%   METHODS        %%%
%%%%%%%%%%%%%%%%%%%%%%%%

\emph{Methods.}
Initial data for the evolutions discussed in this paper correspond to
Brandt-Br\"{u}gmann ``punctures''~\cite{Brandt97b}. The particular
orbital parameters are chosen to be identical to those first evolved
in~\cite{Bruegmann:2003aw}, namely initial separation $L/M=9.32M$,
bare masses $m=0.47656M$ of each BH, and equal and opposite linear
momenta, $p=\pm 0.13808$. These parameters are chosen to approximate a
BH pair in quasi-circular orbit. The angular velocity
$\Omega_0=0.054988$ suggests an orbital timescale of $114M$ for a
perfectly circular orbit. The initial geometry is determined by
numerically solving the constraint equations using the solver of
Ansorg et~al.~\cite{Ansorg:2004ds}.

The binary BH evolution is carried out using the ``BSSN'' formulation
of the Einstein equations~\cite{Nakamura87, Shibata95,
Baumgarte99}, with an implementation described explicitly
in~\cite{Alcubierre02a}. We use a version of the ``1+log'' lapse
and $\Gamma$-driver shifts, given by
Eqs.~(33) (with $f(\alpha) = 2\Psi^m_{BL}/\alpha$)
and (46) (with $F(\alpha)=3/4\,\alpha^p/\Psi^n_{BL}$)
of~\cite{Alcubierre02a}. We generalise the parameter $\eta$ to a
lapse-dependent coefficient $\alpha^q\eta$, where typically $q \in [0,4]$.
This is helpful for long-term evolution of BHs from large initial
separations through merger, and will be discussed in detail elsewhere.

We also dynamically adapt our gauges to the horizon location so as to
approximate co-motion. Individual horizons are located often during
the evolution using the finder described in~\cite{Thornburg2003:AH-finding}.
From the motion of the horizon centroid, the angular and radial
positions and velocities are determined.  Using a damped harmonic
oscillator equation, the acceleration needed to bring the horizon
centroid back to the initial position is determined by
\begin{equation}
  \ddot{\lambda} = -[2 T Q \dot{\lambda}+(\lambda-\lambda^0)]\,/\,T^2.
\label{damping}
\end{equation}
Here $\lambda$ stands for either the current azimuthal angle $\phi$ or
radius $r=\sqrt{x^2+y^2}$ in the orbital plane, and $\lambda^0$ for
its initial value.  $T$ and $Q$ are constants determining the
timescale and damping factor of the harmonic oscillator.  The
correction is added to the time derivative of the shift vector as
\begin{equation}
  \Delta \dot{\beta}^i = (-y,x,0)^i\,\ddot{\theta} - (x,y,0)^{i}\;\ddot{r}/r^0 .
\end{equation}

The punctures are excised using an extension of the ``simple
excision'' techniques which have proven successful in evolving single
BH spacetimes~\cite{Alcubierre00a, Alcubierre01a}. In particular, we
apply the boundary condition to an embedded boundary whose shape is
determined by the horizon location, as described
in~\cite{Alcubierre2003:BBH0-excision,
Alcubierre2003:pre-ISCO-coalescence-times}.

Spatial differentiation is performed via straightforward
finite-differencing, incorporating nested mesh-refined grids with the
highest resolution concentrated in the neighbourhood of the individual
horizons. The mesh refinement is implemented via the Carpet
driver~\cite{Schnetter-etal-03b} for Cactus. The evolutions
carried out in this paper made use of 8 levels of fixed 2:1 refinement. We
fix the regions of increased resolution around the initial BH
locations. Because we are making use of a BH-adapted gauge, the
individual horizons remain on the fine grids throughout the evolution
without requiring moving grids or excised regions. We have used
finest grid resolutions of $h=0.025M, 0.02M, 0.018M, 0.015M$ and
$0.0125M$, with an outer boundary at $96M$ in all cases. Our overall
finite differencing accuracy is second order in space and time.

As a diagnostic of the dynamics of the individual BHs, we
measure the proper distance within a slice between the pair of
apparent horizons (AHs). This is calculated by shooting spacelike
geodesics from the origin (taking advantage of the spacetime symmetry)
to one of the horizon surfaces~\cite{Koppitz2004:PhD}. As
this measure of distance is within each slice, it does depend on the
particular lapse condition (as will be seen below), but still provides
useful information about the near-zone BH motion in the given
slicing.

%%%%%%%%%%%%%%%%%%%%%%%%
%%%   RESULTS        %%%
%%%%%%%%%%%%%%%%%%%%%%%%

\begin{figure}
\includegraphics*[width=18pc,height=14pc]{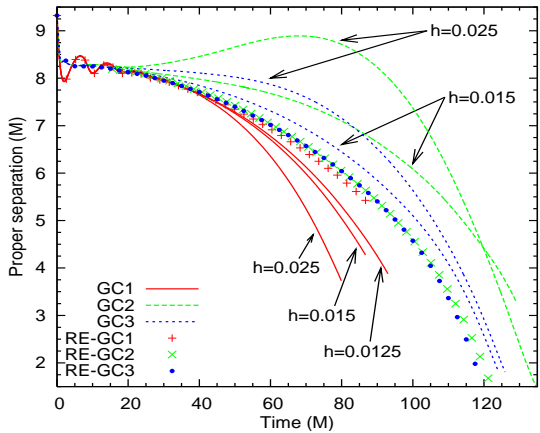}
\caption{The minimal proper distance between individual AHs as a
	function of time. Lines show two representative resolutions,
	$h=0.025M$ and $h=0.015M$ for each gauge choice $GC1$, $GC2$,
	and $GC3$ discussed in the text, and an $h=0.0125$ evolution
	for $GC1$.  Points show Richardson extrapolations (RE)
	for resolutions
	$h \,{\in}\, \{0.018,0.015,0.0125\}M$ for $GC1$,
	$h \,{\in}\, \{0.020,0.018,0.015\}M$ for $GC2$, and
	$h \,{\in}\, \{0.025,0.020,0.015\}M$ for $GC3$.
	}
\label{fig:proper_distance}
\end{figure}

\emph{Results.}
Evolutions were carried out for a number of resolutions and gauge
parameters. A first observation, important in interpreting the results
of~\cite{Alcubierre2003:pre-ISCO-coalescence-times,Bruegmann:2003aw},
is that the infall coordinate trajectory
at a given resolution depends crucially on the gauge choice.  In
Fig.~\ref{fig:proper_distance} the results of evolutions of three
choices of gauge parameters (introduced above) are displayed.
The first, which we label ``$GC1$'', sets
$(m,n,p,q,\eta,T,Q)=(0,2,1,1,4,5,1)$.
The second gauge choice, ``$GC2$'', sets parameters to $(4,2,1,4,2,5,1)$.
The third gauge choice, ``$GC3$'', is described below.
Plotted are the proper separations of the
AHs within the slice versus the coordinate time.

We note that for the $GC1$ evolutions at a grid spacing of $h=0.025M$,
the BHs fall together rather rapidly. By coordinate time $t=75M$,
the proper distance between the AHs is down to $L/M=4.5$.  Increasing
the resolution to $h=0.020M$, $h=0.018M$, and $h=0.015M$ there is a
trend towards longer evolution times before reaching the same
separation, though the convergence is slow. For the highest resolution
which we have evolved, $h=0.0125$, the infall to the same separation
is delayed to only $t=88M$.

The evolutions resulting from gauge choice $GC2$ have an entirely
different character. At the lowest resolution, with $h=0.025M$, we
find that not only does the system evolve well beyond an orbital time
period before a common AH appears, the orbit appears to be elliptical,
with proper separation first increasing with time, and then falling
back, reaching a separation $L/M=4.5$ only after $t=119M$.
This behaviour is qualitatively similar to what was reported for the
same initial data set in~\cite{Bruegmann:2003aw}. Increasing the
resolution as above, however, we find a faster coalescence as
resolution is increased, and the apparent elliptical nature of the
orbit disappears.

Importantly, the two families of evolutions corresponding to $GC1$ and
$GC2$ do show a tendency towards each other as resolution is
increased. In fact, a 3-term Richardson extrapolation (eliminating
2nd and 3rd~order error terms) of the computed
trajectories show that the two families converge to a common
result in the continuum limit.  Fig.~\ref{fig:proper_distance}.
also shows the results of the extrapolation for both sets of evolutions
at the 3~highest resolutions where we had sufficient data available.
(Using other sets of 3 resolutions gives very similar results.)

The differences between $GC1$ and $GC2$ results may at first seem to
be the expected result of the variation in slicing condition between
the two classes of evolutions. However, the fact that the results
converge to the same coordinate trajectory in the continuum limit suggests a
somewhat different explanation, namely that the finite difference
error inherent in the evolution (particularly in the lapse) can be
strongly influenced by the gauge choice.

The Richardson extrapolation provides an indication of the accuracy
of the evolutions at a given resolution, or conversely, the resolution
required to obtain a result of a given accuracy.  The Richardson-extrapolated
coordinate trajectory seems robust, but we note that the highest resolution
(h=0.0125M) $GC2$ simulations which we have carried out still
represent a 22\% error in separation at $t=100M$ compared to the
extrapolated result. The sensitive dependence of accuracy on gauge
suggests that ``ideal'' gauge conditions may be able to improve the
accuracy greatly at a given resolution.

Based on results of the previous experiments, a third gauge choice
parameters was found to demonstrate this point.  Under the label
$GC3$, evolutions of the same data were carried out using gauge
parameters (4,2,1,4,2,5,2.6), i.e.\ the same as for $GC2$ except for the
drift-correction damping parameter.  The $GC3$ results are also plotted
in Fig.~1. Once again, we find the proper separation measure to be
convergent with increasing resolution, and further that the Richardson
extrapolation lies on top of the curves predicted by the previous
evolutions. However, in this case the highest resolution attempted,
$h=0.015M$, lies much closer to the limiting trajectory.

\begin{figure}
\includegraphics*[width=18pc,height=10pc]{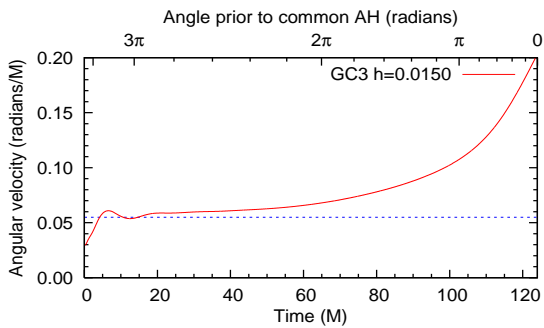}
\caption{%%%
	Angular motion of individual AHs as a function of
	time (lower scale) and angle (upper scale, in radians measured
	backwards along the orbit from the first appearance of a common AH).
	The curve shows the angular velocity $\Omega$ of the
	AH centroid for $GC3$, as measured by integrating the
	corrections applied to the gauge via~\protect\eqref{damping}. The
	horizontal line corresponds to the $\Omega_0=0.055$ estimated
	for the initial data.%%%
	}
\label{fig:omega}
\end{figure}

\begin{figure}
\begin{center}
\includegraphics*[width=25pc]{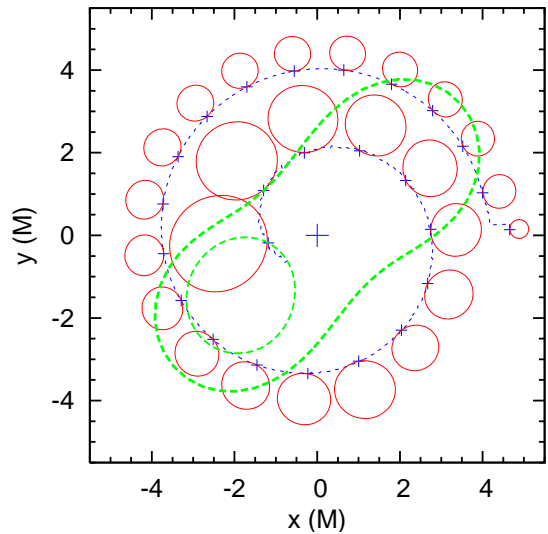}
\vspace{-10mm}
\end{center}
\caption{%%%
	Schematic showing the motion of one of the BHs with time, for the
	$GC3$ gauge choice at the highest resolution $h=0.015M$. At
	intervals of $t=5M$, the AH cross-sections in the $xy$~plane are
	plotted by transforming the co-rotating coordinate system by the
	specified angle and distance. The apparent growth of the AHs with
	time is a non-physical coordinate effect. The first appearance of
	a common AH at $t=124M$, and corresponding final single AH, are
	shown superposed on the figure as dotted lines.%%%
	}
\label{fig:horizons}
\end{figure}

Fig.~\ref{fig:omega} plots
the evolution of the individual BHs' angular velocity, as
measured by the AH centroid's drift, for $GC3$ at our highest
evolved resolution.  From our experience with more closely separated
binaries, the initial shift profile corresponded to a
quasi-rigid rotation with $\Omega=0.3$, approximately $60\%$ of the
nominal $\Omega_0=0.055$ of the initial data~\cite{Tichy:2003qi}.
The co-rotation shift adjustment~\eqref{damping} quickly
raises the effective angular velocity of the coordinate system
up to a value close to the initial data prediction $\Omega_0$.
As expected, this angular velocity then gradually increases
as the BHs spiral together.

Fig.~\ref{fig:horizons} plots the overall motions of the AHs
for the same $GC3$ case as fig.~\ref{fig:omega}.
The individual horizon shapes are shown at intervals of $5M$,
transformed according to the radial and angular motion determined
above.  The initial rapid dip in separation
from $9.32M$ to $8.3M$ in~Fig.~\ref{fig:proper_distance} can now be seen
to be largely a result of the initial coordinate expansion of the
individual horizons, due to our choice of zero radial shift in the
initial data.

A common AH is first detected at $t=124M$, by which time an angular
displacement of $10.2$ radians, or $1.6$ revolutions, has taken
place. The common horizon is found via the method of pre-tracking, in
which a family of surfaces with smallest possible generalised
expansion is followed to provide an estimate which converges on the
first common AH~\cite{Schnetter04}. The listed time is
expected to correspond to the first genuine appearance of a common AH,
independent of the search algorithm. Counting backwards from the
appearance of the common horizon, the duration of the last orbit is
approximately $59M$. It is interesting to note that for the cases
where a common horizon is found, it occurs when the proper separation
is approximately $1.8M$.  Our Richardson-extrapolated $GC1$ and $GC2$
evolutions reach this same separation less than $5M$ earlier than
this $GC3$ evolution.

Given the demonstrated resolution dependence of these results, it
is likely that these results will be subject to some modification as
more accurate evolutions become available.
It should also be recalled that a merged event
horizon within the slice will have formed earlier than the common
AH. Experience with closely separated binaries suggests this is
typically not more than about $5M$ before the appearance of the first
common AH, but studies of the event horizon evolution for these
spacetimes will need to be the subject of future studies.  We also
note that the notion of an ``orbit'' for closely separated BHs is an
intrinsically gauge-dependent quantity -- for
example, a sufficiently small lapse in the region of the horizons
could be used to indefinitely delay merger. The slicings used here are
quite similar in profile to maximal slices, and thus not atypical for
numerical relativity simulations.

As a final point, we note that a number of measures of accuracy have
been monitored during the course of these evolutions. In particular,
the constraint violation is found to remain below a value of $0.05$ at
all points outside the BH horizons and away from the boundaries for the
duration of the runs. The AH masses of the individual BHs were measured and
found to be essentially constant, convergent and accurate even for the low
resolution runs. We have also compared binaries with closer
separation, such as the ``QC-0'' data evolved by a number of
groups~\cite{Alcubierre2003:pre-ISCO-coalescence-times,Campanelli05a,Baker05a}
and find excellent agreement with published results.

%%%%%%%%%%%%%%%%%%%%%%%%
%%% CONCLUSIONS      %%%
%%%%%%%%%%%%%%%%%%%%%%%%

{\em Conclusions.}
We have carried out evolutions of BBH configurations, from initial
data in quasi-circular orbit, through plunge, to the formation of a
common horizon. The dynamics exhibit a number of interesting
properties.

Using a measure of angular velocity based on the AH motion, we have
found that the black holes remain separated for more than $1.5$
revolutions. This measure initially yields an angular velocity
which is consistent with the initial data model.
The most accurate estimate of the duration of the last complete orbit before
formation of the first common AH was $59M$. As might be expected for bodies
falling towards each other, this orbital period is considerably less than is
predicted by the initial angular velocity, $114M$.  The results have
been carried out at resolutions much higher than similar studies to
date, and exhibit good numerical consistency under a range of
resolutions and for a variety of gauge parameter choices.

By Richardson-extrapolating the black hole trajectories we derived
what appears to be a robust extimate of their continuum limit,
which proved to be almost independent of gauge choice, at least
within the family of gauges considered here.

These evolutions required extremely high resolution to attain good
accuracy.  Insufficient resolution can result in very different
predictions for the orbital trajectories and the period of the
final orbit.  This period directly influences the phase of
asymptotically observed wave forms at their strongest point,
and thus is crucial to reproduce accurately.

We demonstrated that the gauge choice can feed directly into the
numerical accuracy of the solution, due to unwanted dynamics in the
evolution variables caused by gauge effects. A preferred gauge was
found to significantly improve the accuracy at a given resolution.
A more detailed exposition of the gauge conditions used here, applied
to a sequence of initial data configurations, will be the subject of
upcoming studies.

%%%%%%%%%%%%%%%%%%%%%%%%%%%
%%%   ACKNOWLEDGMENTS   %%%
%%%%%%%%%%%%%%%%%%%%%%%%%%%

\emph{Acknowledgments.}
We thank B.~Br\"{u}gmann for many
discussions and for sharing details of his codes,
M.~Ansorg for providing the initial data solver,
and
L.~Rezzolla and M.~Alcubierre for many discussions and suggestions.
We have used computing time allocations at the
AEI, CCT, LRZ, NCSA, NERSC, PSC and RZG.
We use the Cactus code infrastructure with a number of locally
developed thorns.
This work was supported in part by
DFG grant SFB TR/7 ``Gravitational Wave Astronomy'',
by the Center for Computation \& Technology at LSU,
and
the Center for Gravitational Wave Physics at PSU.

%%%%%%%%%%%%%%%%%%%%%%
%%%   REFERENCES   %%%
%%%%%%%%%%%%%%%%%%%%%%

\bibliographystyle{bibtex/apsrev-nourl-manyauthors}
\bibliography{bibtex/references}

%%%%%%%%%%%%%%%
%%%   END   %%%
%%%%%%%%%%%%%%%

\end{document}